\documentclass[aps,prb,twocolumn,superscriptaddress,floatfix,nobibnotes]{revtex4-1}

\usepackage[utf8x]{inputenc}
\usepackage{amsmath,subfigure}
\usepackage[pdftex]{graphicx}
\usepackage{textcomp}
\usepackage[version=3]{mhchem} % chemical expressions \ce{H2O}

\usepackage{float} %define float-environment for copyright-notice on 1st page 
\floatstyle{boxed}

\usepackage{hyperref} %always last usepackage!!! See: http://en.wikibooks.org/wiki/LaTeX/Packages/Hyperref
\hypersetup{
    colorlinks=true,       % false: boxed links; true: colored links
    linkcolor=red,          % color of internal links
    citecolor=blue,      % color of links to bibliography
    filecolor=magenta,  % color of file links
    urlcolor=blue,           % color of external links
    urlbordercolor={0 1 1}}	% color of frame around URL links

\newcommand{\degree}{$^{\circ}$}

\begin{document}

\newfloat{copyrightfloat}{thp}{lop}
\begin{copyrightfloat}
\raggedright
The peer reviewed version of the following article has been published in final form at Phys. Chem. Chem. Phys., 2011, 13, 16579, doi: \href{http://dx.doi.org/10.1039/c1cp21607d}{10.1039/c1cp21607d}.
\end{copyrightfloat}

%\preprint{}

\title{Observation of bi-polarons in blends of conjugated copolymers and fullerene derivatives}

\author{Tom J. \surname{Savenije}}\email{T.J.Savenije@tudelft.nl}
\affiliation{Experimental Physics VI, University of W\"urzburg, Am Hubland, 97074 W\"urzburg, Germany}
\affiliation{Opto-electronic Materials, Department of Chemical Engineering, Delft University of Technology, Julianalaan 136, 2628 BL, Delft, The Netherlands}
\author{Andreas \surname{Sperlich}}
\author{Hannes \surname{Kraus}}
\affiliation{Experimental Physics VI, University of W\"urzburg, Am Hubland, 97074 W\"urzburg, Germany}
\author{Oleg G. \surname{Poluektov}}
\affiliation{Chemical Sciences and Engineering Division, Argonne National Laboratory, 9700 Cass Avenue, Argonne, Illinois 60439}
\author{Martin \surname{Heeney}}
\affiliation{Department of Chemistry, Imperial College London, London, SW7 2AZ, UK}
\author{Vladimir \surname{Dyakonov}}
\affiliation{Experimental Physics VI, University of W\"urzburg, Am Hubland, 97074 W\"urzburg, Germany}
\affiliation{Bavarian Centre for Applied Energy Research e.V. (ZAE Bayern), 97074 W\"urzburg, Germany}

%\date{April 14, 2010}

\begin{abstract}
From a fundamental and application point of view it is of importance to understand how charge carrier generation and transport in a conjugated polymer (CP):fullerene blend are affected by the blend morphology. In this work light-induced electron spin resonance (LESR) spectra and transient ESR response signals are recorded on non-annealed and annealed blend layers consisting of alkyl substituted thieno[3,2-b]thiophene copolymers (pATBT) and the soluble fullerene derivative [6,6]-phenyl-C61-butyric acid methyl ester (PCBM) at temperatures ranging from 10 to 180~K. Annealing of the blend sample leads to a reduction of the steady state concentration of light-induced PCBM anions within the blend at low temperatures (T = 10~K) and continuous illumination. This is explained on the basis of the reducing interfacial area of the blend composite on annealing, and the high activation energy for electron diffusion in PCBM blends leading to trapped electrons near the interface with the CP. As a consequence, these trapped electrons block consecutive electron transfer from an exciton on a CP to the PCBM domain, resulting in a relatively low concentration charge carriers in the annealed blend. Analysis of the transient ESR data allows us to conclude that in annealed samples diamagnetic bi-polaronic states on the CPs are generated at low temperature. The formation of these states is related to the generation and interaction of multiple positive polarons in the large crystalline polymer domains present in the annealed sample.
\end{abstract}


\maketitle

\section{Introduction}
Soluble conjugated polymers (CPs) are intensively investigated for application in molecular optoelectronics as they can be processed by low cost solution-based techniques such as ink-jet printing or spin coating. The possibility to realize efficient light-induced generation of free charges in conjugated polymer:buckminster fullerene composites is of great promise for use in so-called bulk heterojunction (BHJ) solar cells~\cite{Hoppe:2004iq,Gunes:2007eo,Blom:2007cf,Brabec:2008tb,Clarke:2010gb,Deibel:2010do}. BHJ solar cells based on CPs and fullerene derivatives reach photovoltaic power conversion efficiencies of more than 8\%~\cite{Green:2011hq}. Due to the low dielectric constant of these polymers, absorption of a photon leads predominantly to formation of a strongly bound electron--hole pair denoted as exciton. In order to convert the exciton efficiently into charge carriers, the CP is blended with a strong electron acceptor, $e.g.$ [6,6]-phenyl-C61-butyric acid methyl ester (PCBM)~\cite{Gregg:2003iv}. After electron transfer from the polymer to PCBM a charge transfer state is formed. To escape from recombination the electron and hole must consecutively dissociate to form free charge carriers, so that collection by the electrodes can occur.

For blends containing crystalline polymer or fullerene domains it has been suggested that in the absence of electrodes light-induced charge carriers populate the crystalline domains~\cite{Clarke:2008gc,Savenije:2004ec,Savenije:2005fv}. Generation of multiple radical cations on CPs might lead to the formation of singlet bi-polaronic states as has been reported to occur both experimentally~\cite{Nowak:1987ia,Fichou:1990jx,vanHaare:1998cf,Behrends:2010hj,Marumoto:2006jq} and theoretically~\cite{Lima:2006kc}. Such a bi-polaron consists of two similar spin 1/2 charges associated with a strong lattice distortion \cite{Bredas:1985dk}.

From fundamental and application point of view it is of importance to understand how the nanomorphology of the blend affects the above processes of blend layers. For blends of poly(alkyl-thiophene) analogues with PCBM much information on the	photophysical processes	is obtained from (time-resolved) optical, microwave and terahertz studies~\cite{Savenije:2005fv,Clarke:2009bl,Marsh:2010gw,Grzegorczyk:2010ep}. However, demonstration of $e.g.$ the formation of bi-polarons using these techniques is not conclusive~\cite{vanHaare:1998cf}. To elucidate whether bi-polarons are formed in bulk heterojunction blends we applied light-induced electron spin resonance techniques (LESR). LESR allows detailed study of the formation and decay of paramagnetic species such as positive and negative polarons and triplets. Investigations using LESR on the formation of positive polarons on the CPs and fullerene anions have been carried out for a number of different blends consisting of a fullerene or fullerene derivative mixed with PPV~\cite{Dyakonov:1999ub,Ceuster:2001hc,Aguirre:2008bw,Schultz:2001fn}, with oligo alkylthiophenes~\cite{Janssen:1995gi} or poly(3-alkylthiophene) (P3AT)~\cite{Aguirre:2008bw,Sensfuss:2003fy,Kuroda:2007tk,Marumoto:2004vu,Marumoto:2003gy,Krinichnyi:2007du,Krinichnyi:2009eq} and with polyfluorene derivatives~\cite{Marumoto:2009gb}. From microwave intensity dependent measurements of the partially overlapping LESR signals observed in thiophene related:PCBM blends, it was inferred that mainly weak-interacting charge carriers are formed~\cite{Dyakonov:1999ub}.

A complicating factor for the interpretation of LESR spectra from $e.g.$ P3HT:PCBM blends arises from the presence of a persistent ESR signal, which is attributed to the long-lived population of deep traps on the polymer~\cite{Smilowitz:1993kc}. Substituted poly(thieno[3,2-$b$]thiophenes) copolymers have a $ca.$ 0.3~eV higher oxidation potential (higher ionisation potential) than P3ATs, which make them more stable for oxidation~\cite{Mcculloch:2009cn}. For this reason it is expected that the persistent ESR signal of thienothiophenes blends will be much smaller than those of P3ATs blends. This work aims at describing the effect of the morphology on the photophysical products generated on illumination of blends consisting of pATBT as electron donor and PCBM as electron acceptor in a 1 to 1 weight ratio (see Fig. \ref{figure1} for molecular structure). In particular we study the formation of diamagnetic bi-polaronic species in these blends. The morphology of the blend is manipulated by annealing as has been studied previously in detail~\cite{Grzegorczyk:2008dt}. In that work it was found that spin-coated pATBT:PCBM blends from dichlorobenzene show randomly sized features of 5--20~nm as determined from TEM images. This phase segregation could be increased by additional thermal annealing yielding large 50--100~nm sized PCBM domains, with a crystalline nature as concluded from XRD measurements. The nanomorphology of the annealed blends shows clear similarities with P3HT:PCBM blends~\cite{Savenije:2005fv,Grzegorczyk:2010ep,Grzegorczyk:2008dt,Yang:2005bt}. Here we present LESR spectra recorded at temperatures varying from 10~K up to 180~K under continuous illumination at 488~nm. From the spectra of the annealed pATBT:PCBM blend (pATBT:PCBM$_{\text{AN}}$) it is inferred that at 10~K diamagnetic bi-polarons are formed, while this process does not occur in the non-annealed sample (pATBT:PCBM$_{\text{NA}}$). As observed from time-dependent LESR measurements recorded at 180~K, the decay of charges is much slower for the pATBT:PCBM$_{\text{AN}}$ indicating that annealing positively effects the formation of long-lived light-induced charge carriers of importance for photovoltaic applications.

\section{Experimental}

The polymer was synthesized as previously reported~\cite{Mcculloch:2009cn}. Molecular weight determinations were carried out in chlorobenzene at 60~\degree C on an Agilent 1100 series HPLC using two Polymer Laboratories mixed B columns in series and were calibrated against narrow weight PL polystyrene calibration standards. The molecular weights for pATBT-C$_{\text{16}}$ were: $M_{\text{n}}$ 36 300 g mol$^{-1}$ and $M_{\text{w}}$ 86 600 g mol$^{-1}$.

Solution mixtures of the copolymer and PCBM containing 8 mg ml$^{-1}$ polymer were prepared in a nitrogen glovebox by dissolving the compounds in 1~:~1 weight ratio in 1,2-dichlorobenzene (ODCB) and stirring overnight at a temperature of 50~\degree C. Typically 150 microlitres of the blend solution was transferred into a 4~mm diameter quartz ESR tube. The tube was connected to a vacuum line to evaporate the solvent, while keeping the sample at room temperature. After reaching a pressure below 10$^{-2}$~mbar the tubes were sealed using a blow torch. To anneal, samples were heated on a hot plate for 25 minutes at 130~\degree C. LESR spectra were recorded using a modified X-band spectrometer (Bruker, ER200D) with an optically accessible microwave cavity (Bruker, ER4104OR). The g-factor was calibrated for each measurement using a high precision NMR-Gaussmeter (Bruker, ER035M) and a microwave frequency counter EIP 28b. During the transient LESR measurements the spectral position (g-factor) was stabilized by a field-frequency-lock unit. For each temperature dependent measurements samples were cooled from room temperature down to the desired temperature using a He flow cryostat (Oxford ESR900), followed by recording the spectrum in the dark. For illumination an Ar Ion laser was used (Melles Griot, 43SE) yielding 488~nm with an intensity of 12~mW unless otherwise stated. At this wavelength the polymer absorbs a large fraction of the incident light. First derivative spectra were recorded using a 100~kHz magnetic field modulation of 2~Gauss.

\section{Results and discussion}

\begin{figure}[ht]
 \centering
	\includegraphics[width=.72\columnwidth]{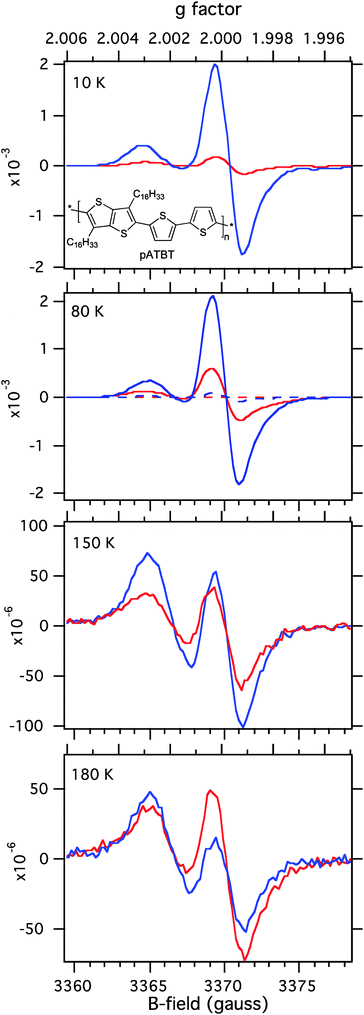}
	\caption{LESR spectra of non-annealed (blue) and annealed (red) 1 : 1 pATBT : PCBM blends recorded on continuous excitation at 488~nm using a microwave power of 0.06, 2, 2 and 20~mW at 10, 80, 150 and 180~K, respectively. At 80~K also dark spectra (dashed lines) are shown. Inset shows the molecular structure of pATBT.}
	\label{figure1}
\end{figure}

\begin{figure}[ht]
 \centering
	\includegraphics[width=\columnwidth]{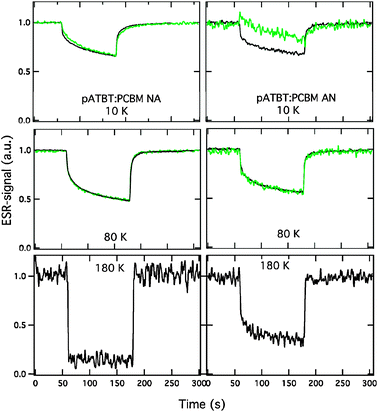}
	\caption{Normalised transient response of the ESR signal for both samples recorded at a g-factor of 2.0035 (positive polaron, green) and at g = 1.9995 (PCBM anion, black). The laser excitation at 488~nm was blocked after 60~s and switched on after 180~s. A microwave power of 0.06 mW was used for measurements at 10~K, 2~mW at 80~K and 60~mW at 180~K. For the latter temperature only the ESR signal at a g-factor of 1.9995 is shown.}
	\label{figure2}
\end{figure}

Samples containing only the pure polymers were tested for their dark and light-induced ESR signal. In contrast to typical ESR signals obtained for different batches of P3ATs~\cite{Smilowitz:1993kc} the poly-thienothiophene copolymer, pATBT, shows only very modest dark and light-induced ESR signals (data not shown). As reported in the literature for pure PCBM no dark signal is found, while only a weak LESR signal is observed~\cite{Dyakonov:1999ub}. For non-annealed and annealed blend samples, the dark first derivative ESR spectra at 80~K were recorded, showing only a small signal as shown in Fig.~\ref{figure1}. On illumination a dramatic enhancement ($>$10~times) of the ESR signal is detected for both blends. The spectra show two overlapping LESR signals with g-factors close to 2.002 and 2.000 indicative for the presence of radical cation on the polymer, denoted hereafter as polaron, and of a PCBM anion, respectively. Similar g-factor values were reported for P3AT:PCBM blends~\cite{Aguirre:2008bw,Krinichnyi:2009eq,Marumoto:2002ts}. To investigate the dependence of both LESR signals on the microwave intensity, LESR spectra were recorded at different microwave powers (see Supporting Information A for measurements at 80 K). Independent of the blend system the ESR signal at g~=~2.002 corresponding to the polaron goes through a maximum on increasing microwave power, indicating that saturation occurs. For the PCBM anion almost no saturation occurs, which has been attributed to the much shorter spin relaxation times for the PCBM anion. Similar dependencies for the polaron and PCBM anion are reported for P3AT:PCBM blends~\cite{Marumoto:2002ts}. For the spectra shown in Fig.~\ref{figure1} saturation of LESR signals was suppressed by choosing a sufficiently low microwave power at each temperature.

Due to different spin lattice relaxation times for the electron on the PCBM and the hole on the polymer, the sizes for their (integrated) ESR signals are different~\cite{Weil:2008tf}. Hence generation of an equivalent number of holes and electrons does not (have to) yield similar ESR intensities. Since the spin--lattice and spin--spin relaxation times are also a function of temperature, a straightforward comparison of the LESR signal sizes can hence only be made for measurements recorded at the same temperature and the same radical species. The ratio between the signal sizes at g~=~1.9995 corresponding to the PCBM anion of the pATBT:PCBM$_{\text{NA}}$ and pATBT:PCBM$_{\text{AN}}$ changes from 11 at 10~K, to 3.8 at 80~K, 1.6 at 150~K to a ratio of 0.68 at 180~K (Fig.~\ref{figure1}). Since the optical absorption of both samples at the used excitation wavelength is very similar, the large differences found in the LESR signals cannot be ascribed to differences in optical absorption. Dissimilar relaxation times for the signal at g~=~1.9995 of both samples, as has been suggested previously for P3AT:PCBM samples~\cite{Krinichnyi:2009eq} is also not likely, since the dependencies of the LESR signal on the microwave intensity show a similar behaviour. Therefore we conclude that the changing ratio between both samples at g~=~1.9995 is due to differences in the steady state PCBM anions concentrations. Since both samples contain the same compounds, it is concluded that the morphologies of both samples are responsible for the differences in concentration and hence the observed ESR signal height. First, we discuss the much higher signals observed for the pATBT:PCBM$_{\text{NA}}$ sample than for the corresponding annealed sample at 10 and 80~K. This observation is surprising since from transient absorption and microwave conductivity measurements it is inferred that the lifetime for light induced charge carriers is enhanced on annealing for pATBT:PCBM and for P3AT:PCBM blends~\cite{Clarke:2009bl,Grzegorczyk:2008dt}. Yet we explain the present data on the basis of the following model: the more molecularly mixed pATBT:PCBM$_{\text{NA}}$ composite features a large interfacial area between both compounds. On illumination, charge transfer states are formed efficiently at this interface. Secondly, due to the temperature activated hopping-like transport of electrons in PCBM at 80~K and lower~\cite{Grzegorczyk:2010ep,deHaas:2006hg}, electrons are immobile and hence trapped closely to the interface. Finally, it is assumed that recombination of an interfacial charge transfer state is slow at low temperatures. For these reasons a relatively high steady state concentration of charges in the pATBT:PCBM$_{\text{NA}}$ composite at low temperatures is found. In contrast the annealed sample contains large crystalline domains, leading to a much smaller interfacial area within the composite. The presence of electrons residing in the proximity of the interface will retard consecutive electron transfer from an additionally excited polymer to PCBM domain. Hence the LESR signal intensity is reduced in the annealed pATBT:PCBM blend w.r.t. the non-annealed sample. At higher temperatures (150~K and 180~K) the signal sizes at g~=~1.9995 of the non-annealed and annealed sample are similar. At these temperatures the electrons are able to diffuse away from the interface with the polymer resulting in similar concentrations of PCBM anions in both blends.

To investigate the effect of the morphology of the blend on the transient response of the ESR signal, the samples were measured at a fixed magnetic field coinciding with the transition of either the polaron or of the PCBM anion. Typically on cessation of the laser light after 60~seconds of illumination a drop of the LESR signal was observed as shown in Fig. \ref{figure2}. After 180~s the laser was switched on again resulting in a rise of the ESR signal. These experiments could be repeated multiple times at different temperatures and light intensities, showing the excellent stability of the blend samples. The LESR signals were normalized to unity for the duration of the first 60~s. For the measurements carried out at 80~K both samples show on blocking the laser beam a relative fast decay occurring within 10~seconds. Note that the response time of our set-up of 0.1~s, as determined by the time constant of the lock-in amplifier, could significantly affect the temporal shape of the fast decay and rise of the transient ESR signal. The fast decay is followed by a slower decay which extends over hundreds of seconds. For blends of a PPV derivative and PCBM decay constants exceeding hours have been reported~\cite{Schultz:2001fn}. Clearly, at 80~K the transient signal of the polarons and of the PCBM anions overlap perfectly. This overlap indicates the formation (and recombination) of a positive polaron on the polymer and a PCBM anion are coupled processes, $i.e.$ their stoichiometric ratio is 1. The recombination process is given by:

\begin{equation}
\text{pATBT$^+$ + PCBM$^-$ $\rightarrow$ pATBT + PCBM}
\label{process1}
\end{equation}

At 10~K again an apparent overlap between the transient ESR signals of the positive polaron and PCBM anion of the pATBT:PCBM$_{\text{NA}}$ sample is visible (Fig. \ref{figure2}, upper left). Surprisingly at 10~K the signals for the annealed blend do clearly not coincide (Fig. \ref{figure2}, upper right). More specifically, the polaron signal does not exhibit a fast decay on cessation of the light; on the contrary a small rise can be discerned over the first few seconds, followed by a slow decay over hundreds of seconds. The fact that transients do not overlap can only be explained by the presence of polaronic states on the CP, which cannot be observed by ESR $i.e.$ are diamagnetic. A possible explanation might be the formation of bi-polarons in the annealed pATBT:PCBM blend on illumination at 10~K. Recombination of the bi-polaron proceeds $via$ the formation of a polaron given by process \ref{process2}:

\begin{equation}
\text{pATBT$^{2+}$ + PCBM$^-$ $\rightarrow$ pATBT$^+$ + PCBM}
\label{process2}
\end{equation}

followed by process \ref{process1}. The consecutive processes for the decay of the bi-polaron explain the initial small rise observed for the positive polaron on the annealed sample at 10~K.

To investigate the formation of bi-polarons in more detail similar transient ESR responses of the pATBT:PCBM$_{\text{AN}}$ blend at 10~K were recorded using laser intensities varying from 1~mW up to almost 60~mW. (see Supporting Information B) It turns out that the differences between the transient ESR signals of the polaron and of the PCBM anion become more pronounced on increasing the laser intensity. This observation suggests that on higher light intensities the ratio of bi-polarons to polarons in the pATBT:PCBM$_{\text{AN}}$ sample increases.

\begin{figure*}[ht]
 \centering
	\includegraphics[width=1.5\columnwidth]{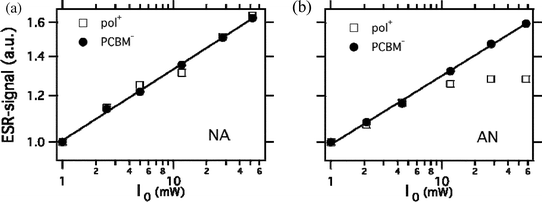}
	\caption{Dependence of the LESR signals of pATBT:PCBM$_{\text{NA}}$ and pATBT:PCBM$_{\text{AN}}$ at 10~K at g~=~1.9995 and g~=~2.0035 as a function of the used laser intensity. The intensities of the polaron ESR signals were adjusted in order to overlay with the values of the PCBM for the lowest light intensity used.}
	\label{figure3}
\end{figure*}

In order to confirm if bi-polarons are formed LESR spectra of pATBT:PCBM$_{\text{AN}}$ using differing light intensities were recorded. From these spectra the signal heights at 1.9995 and at 2.0035 corresponding with the transitions of the PCBM anion and the polaron, respectively, are plotted in Fig. \ref{figure3} as a function of the incident laser power in a double logarithmic representation. The signals were normalized to unity at a laser power of 1~mW. As shown in Fig. \ref{figure3}A for the pATBT:PCBM$_{\text{NA}}$ sample the increase in the signal size on the laser power follows a power law: $Signal = c I_{0}^{\alpha}$. Values for $\alpha$ amount for both charge carriers to 0.121. This similarity indicates again that the increments in concentration of both charge carriers are equal and seem to be coupled. Similar observations have been made previously for other blends~\cite{Dyakonov:1999ub,Schultz:2001fn,Marumoto:2002ts}. For pATBT:PCBM$_{\text{AN}}$ a similar value of a amounting to 0.117 is observed for the PCBM anions (Fig. \ref{figure3}B). However, for the polaron signal a clear deviation is noticeable. This observation again indicates that at high intensities corresponding to high charge carrier concentrations, the LESR intensity saturates, which can be attributed to the fact that in addition to polarons bi-polarons are formed. In contrast, for the pATBT:PCBM$_{\text{NA}}$ this deviation is not observed indicating that the ratio of bi-polarons to polarons is much smaller in comparison to the annealed sample.

For oligo and poly alkyl thiophenes bi-polarons have been observed previously using ESR~\cite{Fichou:1990jx,Horowitz:1994kj}. A possible explanation for the formation of bi-polarons in the annealed sample is related with the presence of the large crystalline domains. On light induced electron transfer from the polymer to the PCBM molecule, the polarons might diffuse towards the crystalline domains and pile up in these domains, leading to the formation of bi-polarons.

In the non-annealed blend the absence of crystalline polymer domains forces the polarons to reside in close proximity of their oppositely charged counterpart. A similar conclusion has been made by Lane $et~al.$ observing bi-polarons in ordered sexithiophenes~\cite{Lane:1998bi}. Interestingly, increasing the temperature from 10 to 80~K leads to disappearance of the bi-polarons as can be concluded from Fig. \ref{figure2}. This would mean that the activation energy for the transition from a bi-polaron to two paramagnetic polarons would be in the order of several meV, which is in agreement with reported values for poly(alkyl thiophene) related materials~\cite{Cik:bn,Kadashchuk:2004if}. Recently, Behrends $et~al.$ reported the formation of positive bi-polarons in a poly(phenylvinylene) derivative:PCBM blend even at room temperature~\cite{Behrends:2010hj}. Additionally, Marumoto $et~al.$ showed, that field-induced positive bi-polarons in P3HT MIS-diodes are detectable~\cite{Marumoto:2006jq}. Possible reasons for explaining the changed temperature behaviour of the bi-polarons include the different materials and measurement techniques resulting in high charge carrier densities.

In Fig. \ref{figure2} the transient ESR signals of PCBM anion (g~=~1.9995) at 180~K are included. For the pATBT:PCBM$_{\text{NA}}$ blend the ESR signal vanishes almost completely on blocking the laser beam. In contrast, for the pATBT:PCBM$_{\text{AN}}$ the signal partially remains, indicating that long-lived charge carriers have been formed under illumination. Apparently, the crystalline domains in the annealed sample partially retard the electron hole recombination process. A reason for this might be that for recombination the charge carriers have to move out of the crystalline domains, which is an energetically uphill process. This is in agreement with previous photoconductance measurements on blends of P3HT:PCBM~\cite{Savenije:2005fv,Quist:2007fl}.

\section{Conclusions}
In this paper we have shown the effects of the nanomorphology within a blend consisting of a thienothiophene copolymer and PCBM, on the processes occurring on optical excitation. From the LESR spectra and transient response signals recorded at various temperatures ranging from 10 to 180~K, the following observations were made. Annealing lowers the steady state concentration of light-induced charge carriers. Secondly, in annealed samples diamagnetic bi-polaronic states on the CPs are generated at 10~K. On increasing the light intensity the ratio of bi-polarons to polarons increases. Finally, only in annealed samples long-lived charge carriers are generated at 180~K

All these observations are explained on the basis of the following reasoning. The interfacial area in a blend composite diminishes substantially on annealing. This issue has been reported previously for similar blend composites using TEM. The temperature activated hopping like transport of electrons in PCBM leads below 80~K to immobile electrons near the interface with the polymer. This assumption is in line with the temperature dependent mobility measurement of excess charges in poly-crystalline PCBM. A major consequence is that these immobile electrons at the interface with a polymer domain block consecutive electron transfer from an exciton on a CP to the PCBM domain. At temperatures above 150~K the excess electrons on PCBM are able to diffuse away from the interface with the polymer resulting in similar concentrations of PCBM anions in both blends (NA and AN). We believe that most of the observed processes are also occurring in blends consisting of other compounds provided that the conjugated polymer forms crystalline domains.

\begin{acknowledgments}
O.P. was supported as part of the ANSER Center, an Energy Frontier Research Center funded by the U.S. Department of Energy, Office of Science, Office of Basic Energy Sciences under Award Number DE-SC0001059. A.S.'s, H.K.'s and V.D.'s work was financially supported by the German Research Council (DFG) under contracts DY18/6-2 and INST 93/557-1.
\end{acknowledgments}

\section*{Supporting Information Available}
This material is available free of charge via the Internet at \href{http://www.rsc.org/suppdata/cp/c1/c1cp21607d/c1cp21607d.pdf}{www.rsc.org}.

\bibliography{hex}

\end{document}